\documentclass{sf2a-conf2016}
\usepackage{graphicx}
\usepackage{siunitx}
\usepackage{hyperref}
\usepackage[]{natbib}  
\usepackage{epstopdf}

\def\BibTeX{{\rm B\kern-.05em{\sc i\kern-.025em b}\kern-.08em
    T\kern-.1667em\lower.7ex\hbox{E}\kern-.125emX}}
\def\arcsec{\hbox{$^{\hbox{\rlap{\hbox{\lower4pt\hbox{$\,\prime\prime$}}
          }\hbox{$\frown$}}}$}}

\def\arcmin{\hbox{$^{\hbox{\rlap{\hbox{\lower4pt\hbox{$\;\prime$}}
          }\hbox{$\frown$}}}$}}
\bibpunct{(}{)}{;}{a}{}{,}  

\begin{document}

\TitreGlobal{SF2A 2016}


\title{The low-frequency radio emission in blazar PKS2155-304}

\runningtitle{The low-frequency radio emission in blazar PKS2155-304}

\author{M. Pandey-Pommier}\address{Univ Lyon, Univ Lyon1, ENS de Lyon, CNRS, Centre de Recherche Astrophysique de Lyon UMR5574, 9 av Charles Andre, 69230, Saint-Genis-Laval, France}
\author{S. Sirothia$^2$}\address{National Centre for Radio Astrophysics - Tata Institute of Fundamental Research, Pune University Campus, Post Bag 3, Ganeshkhind Pune 411007, India}\address{SPT, Cape Town, Square Kilometre Array-South Africa}
\author{P. Chadwick}\address{Department of Physics, Institute for Computational Cosmology, Durham University, South Road, Durham DH1 3LE, United Kingdom}
\author{J.-M. Martin}\address{GEPI, Observatoire Paris CNRS, 5 Pl Jules Janssen, 92195 Meudon, France \& Station de Radioastronomie de Nan\c{c}ay, Observatoire de Paris, 18330 
Nan\c{c}ay, France}
\author{P. Colom$^5$}
\author{W. van Driel$^5$}
\author{F. Combes}\address{LERMA, Observatoire de Paris, CNRS, 61 rue de l'Observatoire, 75014 Paris, France}
\author{P. Kharb$^2$}
\author{P-J. Crespeau$^1$}
\author{J. Richard$^1$}
\author{B. Guiderdoni$^1$}
\setcounter{page}{237}


\maketitle


\begin{abstract}
We report radio imaging and monitoring observations in the frequency range 0.235 - 2.7 GHz during the flaring mode of PKS 2155-304, one of the brightest BL Lac objects. 
The high sensitivity GMRT observations not only reveal extended kpc-scale jet and FRI type lobe morphology in this 
erstwhile `extended-core' blazar but also delineate the morphological details, thanks to its arcsec scale resolution. The radio light curve during the end phase of the 
outburst measured in 2008 shows high variability (8.5$\%$) in the jet emission in the GHz range, compared to the lower core variability (3.2$\%$) seen at the lowest frequencies. 
The excess of flux density with a very steep spectral index in the MHz range supports the presence of extra diffuse emission at low frequencies. The analysis of 
multi wavelength (radio/ optical/ gamma-ray) light curves at different radio frequencies confirms the variability of the core region and agrees with 
the scenario of high energy emission in gamma-rays due to inverse Compton emission from a collimated relativistic plasma jet followed by synchrotron emission in radio. 
Clearly, these results give an interesting insight of the jet emission mechanisms in blazars and highlight the importance of studying such objects with low frequency radio 
interferometers like LOFAR and the SKA and its precursor instruments.
\end{abstract}



\section{Introduction}
Blazars are active galactic nuclei (AGN) powered by accretion onto super massive black holes and mostly associated with BL Lacertae objects (no emission lines) (Urry $\&$ Padovani 1995). 
They show ultra-relativistic outflows and diverse timescale flaring activities from radio (100 MHz) up to Gamma-rays (tens of TeV)(Foschini et al. 2007, 2008). The jets in blazars are 
characterized as compact morphology, flat spectral core regions of high degree of polarization with superluminal speed (Angel $\&$ Stockman 1980). 
It is believed that the recurrent flaring episodes in AGNs may give rise to large-scale diffuse emission around them that may finally terminate into lobes. These radio lobes show
very steep spectra and can be used to probe the AGN duty cycle, their activity history as well as surrounding environmental properties. High sensitivity, arcsec scale resolution
observations at low radio frequencies are needed to detect such diffuse emission in blazars and disentangle the details of their morphology. Kharb et al. (2010) carried out a detailed
search with the Very Large Array (VLA) at 1.4 GHz for the extended jet (kpc-scale) emission in a sample of 135 radio loud blazars from the MOJAVE sample and detected extended
jet emission in the majority of the sources, with only 7$\%$ of the sources exhibiting compact core emission. The kpc-scale jet emission in a large fraction of MOJAVE BL Lac objects showed
radio power and jet morphology typical of FRII type AGNs (contrary to conventional results), while a substantial number of blazars associated with quasars had possessed a radio power 
intermediate between FRIs (core with inflated jet morphology) and FRIIs (core with jet terminating into hotspots)(Kharb et al. 2010, Fanaroff $\&$ Riley 1974). In addition to the 
kpc-scale jet structures, BL Lacs exhibit rapid variable emission (less than an 
hour to several days) from their core region due to relativistic beaming effects of the jets and are therefore frequently monitored during outburst. In this paper we will discuss  
the extended faint emission in PKS 2155-304 discovered in high sensitivity GMRT observations in the MHz range and the flux 
variability in comparison with the single-dish Nan\c{c}ay Radio Telescope (NRT) observations in the GHz range during the end phase of the 2008 flaring episode.

\section{Low frequency diffuse emission in PKS 2155-304}
PKS 2155-304 is one of the most luminous and highly variable BL Lac object located at a redshift of $z$=0.117 (Urry et al. 1993, 1997; Aharonian 2005, 2007, 2009). It was detected as 
a high energy TeV Gamma-ray source in 1996-1997 by the University of Durham Mark 6 Telescope and later confirmed by the High Energy Spectroscopic System (H.E.S.S.) instrument 
(Chadwick et al. 1999, Abramoski et al. 2012, Djannati-Atai et al. 2003). The optical and near-IR data show that the object is hosted by a dominant luminous elliptical surrounded by a 
rich group of galaxies at a similar redshift as PKS 2155-304 (Falomo et al. 1991, 1993). In high resolution ($\sim 0.12''$) NIR $J$ and $K$ band images made with the Multi Conjugate Adaptive
Optics Demonstrator (MAD) system at the ESO (European Southern Observatory) VLT (Very Large Telescope) a search was made for the NIR counterparts in order to investigate the 
properties of the close environment of the source and associated radio counterparts. The observations confirmed the association of a poor galaxy group with the target, but no radio
emission was detected from these galaxies, except for the central host (Liuzzo et al. 2013). The central elliptical host in PKS 2155-305 exhibits slightly resolved morphology at 
radio wavelengths with a core and surrounding halo region extending up to $94''$ or 132 kpc in $8.4'' \times 3.8''$ 
maps with the VLA at 1.5 GHz and a probable very faint extension up to $\sim 200''$ or 375 kpc in size, in low resolution maps ($26'' \times 10''$), which is  
apparently over-resolved or undetected in most of the GHz-range observations (Ulvestad et al. 1986, Laurent-Muehleisen et al. 1993). 
Liuzzo et al. (2013) made VLA radio continuum images from 1.4 to 22.5~GHz and found old radio jet emission at $\sim 20$ kpc from the center of PKS 2155-304 and 
a jet-like structure of $\sim 2$ kpc size in the eastern direction. A knot of 10$''$ in the NW direction was also detected in a high resolution ($\sim 0.6''$) map (Piner et al. 2010). 
The extended radio power measured at 1.5 GHz was log(P$_{ext}$) = 25 W Hz$^{-1}$ (Laurent-Muehleisen et al. 1993).
\begin{figure}[h]
 \centering
 \includegraphics[width=1.0\textwidth, clip]{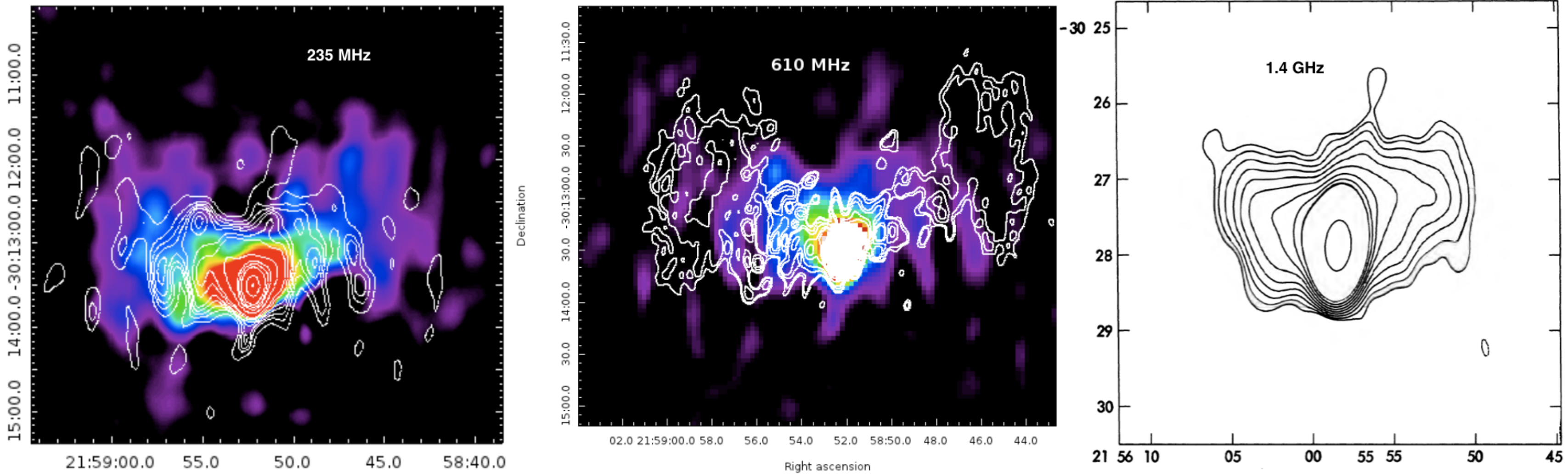}
  \caption{PKS 2155-304: GMRT radio continuum images: (left:) contours (white) at 235 MHz with resolution 21$''$ and rms noise 0.8 mJy/beam overlaid on the 610 MHz color map, (middle:) 
610 MHz contours (white) with resolution 5$''$ and rms noise 0.04 mJy/beam overlaid on the 235 MHz color map and (right:) VLA map at 1420 MHz with resolution 26$'' \times 10''$, extracted 
from Ulvestad et al. (1986).}
 \label{author1:fig1}
\end{figure}

We carried out deep GMRT observations on PKS 2155-304 at 610 and 235 MHz for almost 20 hours and confirmed the
presence of a roughly 4.5 arcmin (575 kpc, for a cosmological model with $\Omega_M$= 0.28 and $h$ = 69.6 km/s/Mpc) sized extended structure down to 235 MHz and resolved 
the structural details-- core, halo, jet and lobe regions (Pandey-Pommier et al. 2015). Parts of diffuse emission of different size scales (few 10s of kpc) had already 
been detected in less sensitive maps previously at different locations in PKS 2155-304
(Beuchert et al. 2010; Liuzzo et al. 2013). However, low frequency GMRT data was able to discover the overall extent of the diffuse emission and resolve the details, 
in order to give a complete morphological profile of PKS 2155-304, for the first time- thanks to its high 
sensitivity and resolution. Further, the spectral index analysis between 610 and 235 MHz GMRT data suggests that the emission is steepest in 
the radio lobe, with $-1 < \alpha < 0$, while the core region shows flatter spectral index with $\alpha = -0.2$, in agreement with VLA observations. 
Detailed analysis of the spectral properties of diffuse jet emission in PKS 2155-304 will be presented in a future publication (Pandey-Pommier et al. 2016).
   
\section{Core variability in PKS 2155-304}
PKS 2155-304 is known to exhibit significant variability, on both long (months) and short (days to hours) time scales from 
radio up to Gamma-ray wavelengths (see Fig. 2, Aharonian et al. 2009). In August 2008, a multi wavelength monitoring campaign was performed on this 
blazar during the end phase of a bright flare detected 
at gamma-rays with H.E.S.S. The RXTE X-ray and H.E.S.S. data showed correlation during the high state of the source, but
no good correlation was seen with the longer radio wavelength data. Nevertheless, a correlation in the radio data from the NRT, 
Australian Telescope Compact Array (ATCA), 
Hartebeesthoek Radio Astronomy Observatory (HartRAO) and optical ($Swift$ UVOT) data was seen following 
the decay in the peak at very high energy (VHE, refer Fig. 2 in Abramowski et al. 2012).
\begin{figure}[h]
 \centering
 \includegraphics[width=1.0\textwidth, clip]{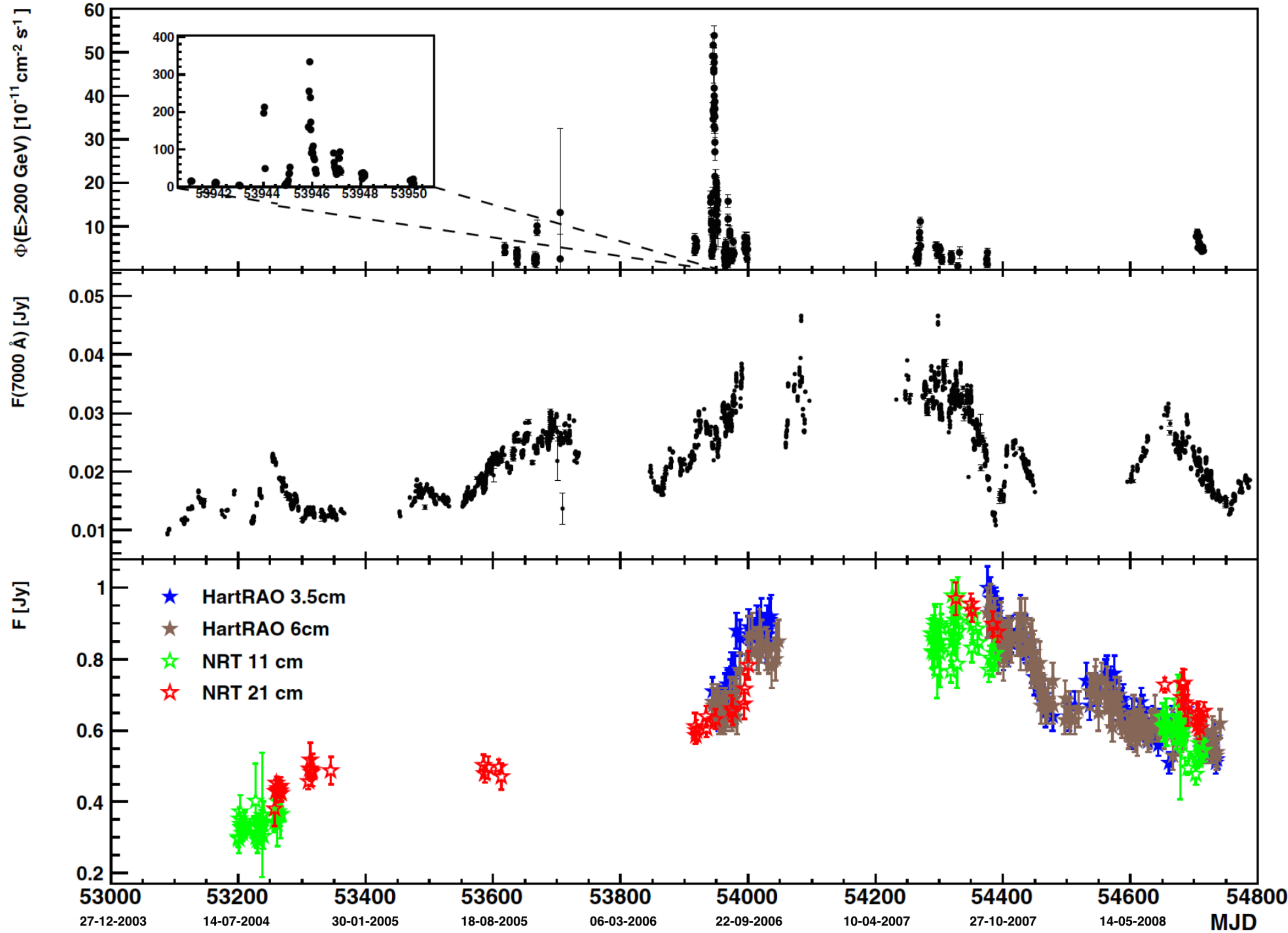}
  \caption{Multi wavelength light curve of PKS 2155-304 during the 2006-2008 flaring episode (extracted from Abramowski et al. 2012).}
\label{author1:fig2}
\end{figure}
 
The radio flux density data collected from 1.4 GHz up to 9 GHz shows that the source was in an active state since 2006 had reached the end of the outburst state by August 2008. 
The average flux density measured by the NRT was 0.49$\pm$0.04 Jy at 2.7 GHz and 0.65$\pm$0.02 Jy at 1.4 GHz, suggesting an excess in 
emission at lower frequency for an assumed radio spectral index of -0.8. The long term variability monitoring multi wavelength data were explained via three different emission 
models, viz., (i) a one-zone synchrotron self 
Compton (SSC) model, where the X-ray and Gamma-ray bumps are generated due to synchrotron and Inverse Compton (IC) emission from a population of relativistic electrons in a 
spherical plasma blob inside the jet medium, (ii) a two-zone and stratified jet SSC model, where high energy emission is generated from the relativistic electrons and positrons 
in an extended, in homogeneous jet and a dense plasma blob traveling along the jet axis. The dense inner blob being more energetic gives rise to VHE flux and rapid variability 
in SSC emission, while the extended jet emission gives rise to underlying continuous outflow. Thus two different jet components emitting synchrotron radiation are considered in this model, and  
(iii) a stratified jet scenario, where only one radiative MHD jet component is considered to be launched by the accretion disk. This jet surrounds a highly relativistic plasma of electron-
positron pairs long its axis. The MHD jet plays the role of a collimator and energy reservoir for the pair plasma being injected at the base of the jet. The plasma is re-accelerated 
along the jet to compensate for synchrotron and IC cooling, which are responsible for observed high enery emission following the SSC scenario. In spite of detailed modeling at high 
frequencies of the variability in PKS2155-304, none of the above scenarios were able to describe the poor correlation
between VHE and the long term radio evolution. The multi wavelength light curve suggested that the rapid activity at high energies on short time scales directly contributes to an
averaged, long-term signature at lower energies (see  Fig. 2, Abramowski et al. 2012).
\begin{figure}[h]
 \centering
 \includegraphics[width=1.0\textwidth, height=0.4\textheight, clip]{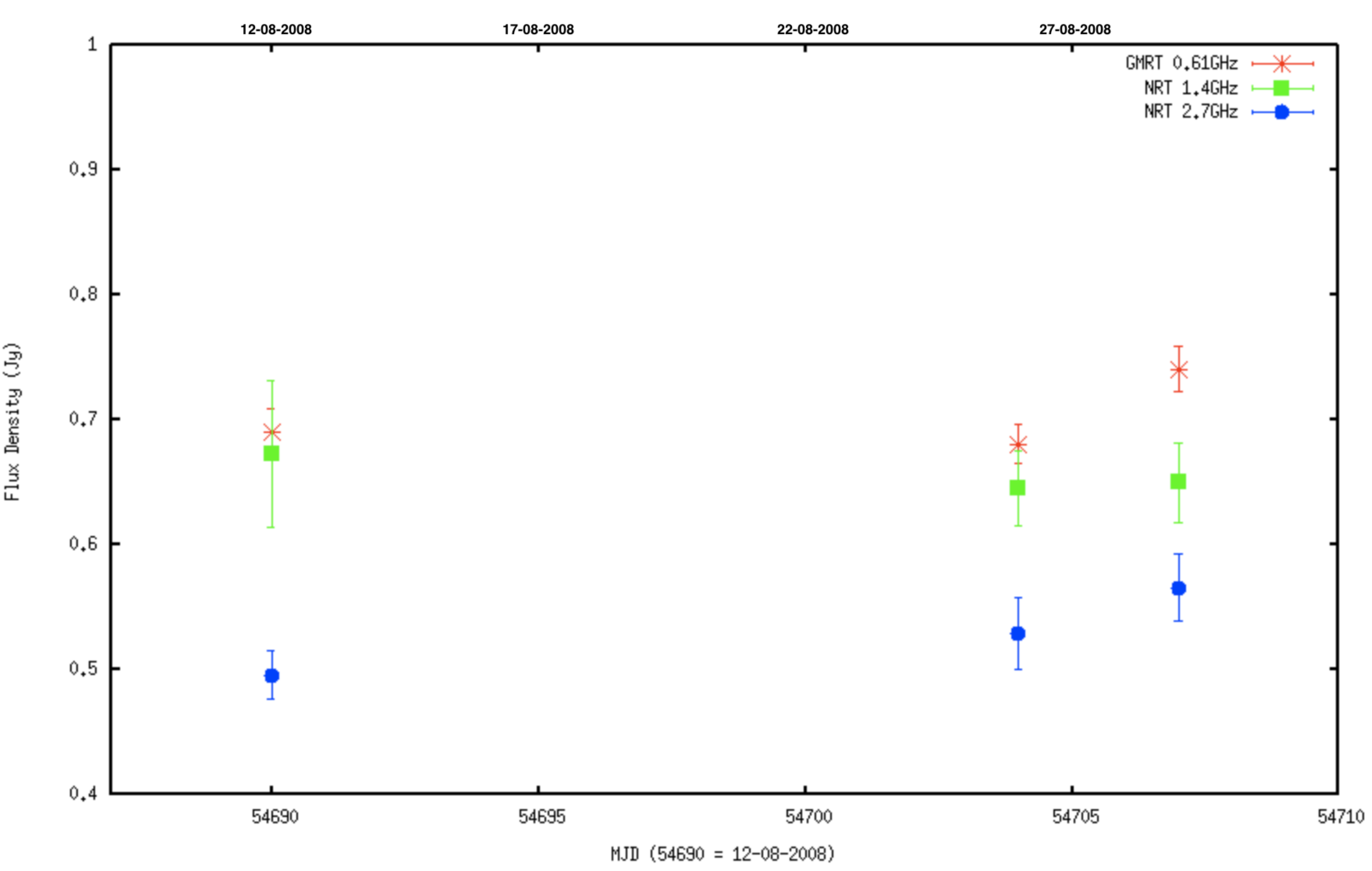}
  \caption{Radio light curve in 2008 during the end of the flaring episode.}
\label{author1:fig2}
\end{figure}

The radio light curve derived from the simultaneous data available at 610 and 235 MHz with the GMRT and at 1.4 GHz and 2.7 GHz with the NRT suggests that 
the low frequency radio emission shows reduced flux variability (3.2$\%$) of the compact core component as compared to that at higher frequencies (8.5$\%$) and that there is an excess 
in the low frequency flux by almost $\sim$34$\%$ (see Fig. 3). At 235 MHz very marginal variability was seen. This clearly supports the scenario that low frequency 
emission is dominated by emission from the non variable extended jet-lobe structure 
that tends to be more luminous at lower frequencies. Furthermore the reduced variability seen at lower frequencies is due to new radio emitting blobs of plasma ejected within 
the jet medium that are initially opaque to radio waves dues to synchrotron self absorption and then slowly expand adiabatically and cool, causing emission in the MHz 
range with a certain delay. The longer cooling times at lower
frequencies leads to a longer timescale of radio emission as compared to very rapid high energy emission flares. The reduced (or diluted) variability in the radio
light curve in the MHz range is due to combined emission from the inner core-jets, superimposed by the dominant low frequency radio emission from the extended outer 
lobe regions, in agreement with above proposed models.

\section{Square Kilometre Array (SKA) survey capabilities}
Radio monitoring of blazars along with multi wavelength data provides an important constraint on their synchrotron self Compton (SSC) emission models as well as kpc-scale jet properties 
(Aharonian et al. 2009, Band and Grindly 1985, Kharb et al. 2016, Pandey-Pommier et al. 2015). Thanks to the upcoming radio facilities with high sensitivity and resolutions like 
the LOw Frequency ARray (LOFAR, 10-240 MHz) and the SKA
(50 MHz - 10 GHz), operating at low frequencies, it will be possible to detect many more such faint jet structures in blazars (see Fig. 4, Kharb et al. 2016).
\begin{figure}[h]
 \centering
 \includegraphics[width=1.0\textwidth, clip]{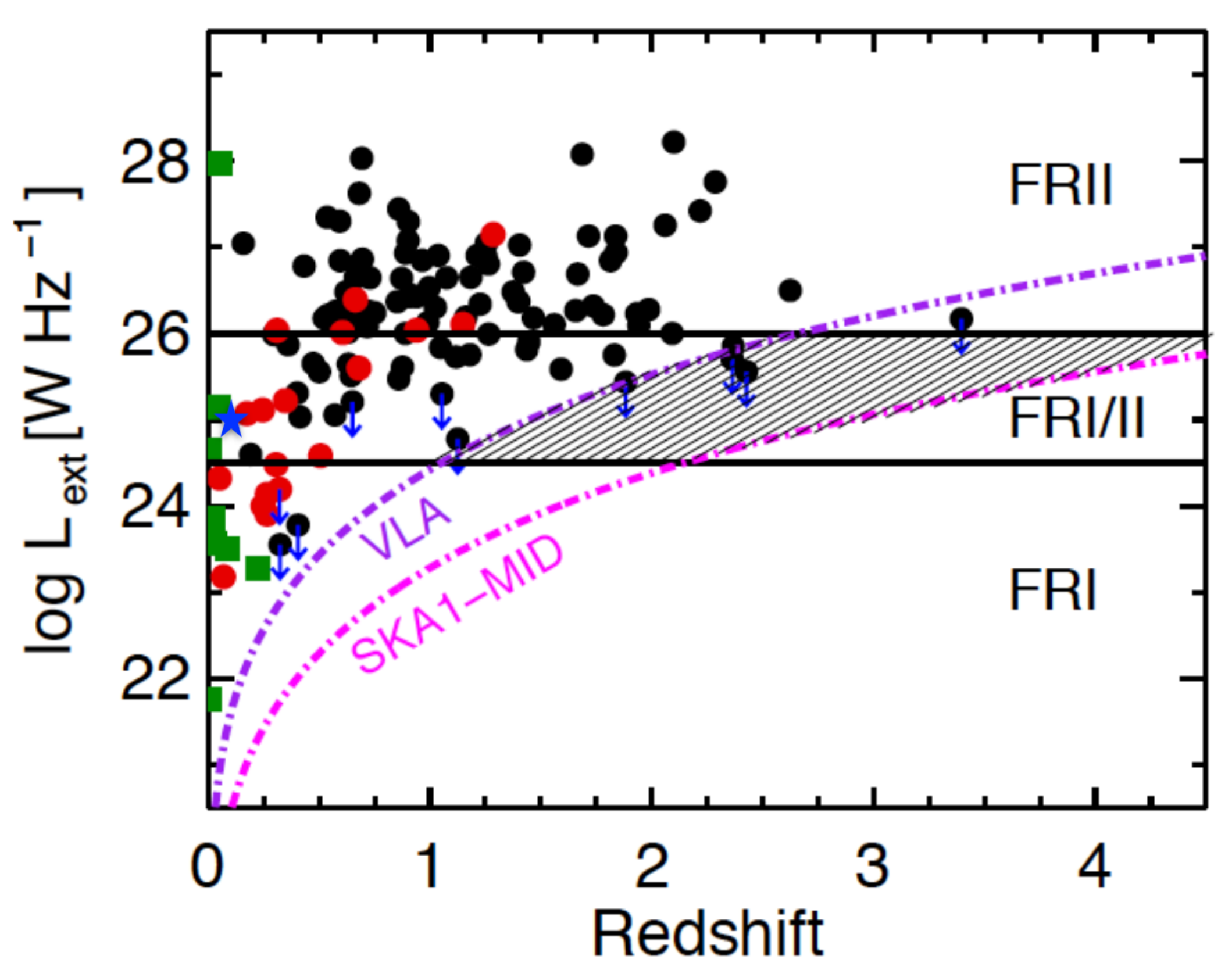}
  \caption{Radio luminosity as a function of redshift for quasars (black circles), blazars (red circle), radio galaxies (green squares) and core-only sources (upper limits with 
downward arrows). 
PKS 2155-304 is indicated as a blue star. The FRI-FRII radio power divide is marked with a solid black line and sensitivity limits for the VLA and the SKA Phase 1 mid-range frequency 
(1.65 to 3.05 GHz) component are shown as dashed lines (adapted from Kharb et al. 2016).} 
 \label{author1:fig1}
\end{figure}

The higher sensitivity of SKA1-MID Band 3 (1.65 to 3.05 GHz) at high frequencies is expected to detect twice as many extended diffuse radio jet emission in blazars than previously 
known with the VLA (Kharb et al. 2016). 
SKA1-MID will offer a factor of 10 to 70 overall increase in the sensitivity (0.7 $\mu$Jy beam$^{-1}$) and will be able to delineate the full extent of diffuse lobe emission in blazars and 
radio galaxies, that were missed out with the previous VLA survey at 1.4 GHz having less sensitivity (10-50 $\mu$Jy beam$^{-1}$). Thus SKA survey will not only discover new 
kpc-scale persistent diffuse jet emission in blazars, but also help to classify their $``$intermediate$"$ or $``$hybrid FRI/II$"$ type morphology. It will also confirm if the 
$``$core-only$"$ blazars are a different population that exhibit episodic jet activity and demonstrate an unresolved core morphology during switched off.     

\section{Conclusions}
We list below the results achieved from studying the non-thermal radio emission from the most luminous blazar PKS 2155-304 down to 0.235 GHz with the GMRT and multi wavelength 
monitoring of flaring activity in 2008. 
\begin{itemize}
\item Low frequency radio emission at 235 MHz is dominated by emission from an extended, diffuse jet-lobe of 575 kpc size discovered with the GMRT;
\item We have confirmed the variability of the source down to 610 MHz during the 2008 flaring episode and its decreased variability at lower frequencies (235 MHz);
\item The compact core is highly variable over time scales of a few hours to days down to 610 MHz and has a flat spectral index;
\item The SKA will play an important role in detecting new populations of FRI/II sources as well as faint blazar jets.
\end{itemize}

\begin{acknowledgements}
We thank the staff of GMRT, who made these observations possible. The GMRT is run by the National Centre for Radio Astrophysics of the Tata Institute of Fundamental 
Research. MP is thankful to the research grant received from the Franco-Indian CEFIPRA organization. 
The Nan\c{c}ay Radio Telescope is operated as part of the Paris Observatory, in association with the CNRS and partially supported by the R\'egion Centre in France. 
\end{acknowledgements}


\bibliographystyle{aa}  
\bibliography{sf2a-template} 
\end{document}